\documentclass[oneside,english]{amsart}
\usepackage[T1]{fontenc}
\usepackage[latin9]{inputenc}
\usepackage{amsthm}
\usepackage{amssymb}

\makeatletter

\providecommand{\tabularnewline}{\\}

\numberwithin{equation}{section} %% Comment out for sequentially-numbered
\numberwithin{figure}{section} %% Comment out for sequentially-numbered

\makeatother

\usepackage{babel}

\begin{document}

\title{MapReduce for Integer Factorization}

\author{Javier Tordable}
\begin{abstract}
Integer factorization is a very hard computational problem. Currently
no efficient algorithm for integer factorization is publicly known.
However, this is an important problem on which it relies the security
of many real world cryptographic systems. 

I present an implementation of a fast factorization algorithm on MapReduce.
MapReduce is a programming model for high performance applications
developed originally at Google. The quadratic sieve algorithm is split
into the different MapReduce phases and compared against a standard
implementation.
\end{abstract}
\maketitle

\section{Introduction}

The security of many cryptographic algorithms relies on the fact that
factoring large integers is a very computationally intensive task.
In particular RSA \cite{1} would be vulnerable if there was an efficient
algorithm to factor semiprimes (products of two primes). This could
have severe consequences, as RSA is one of the most widely used algorithms
in electronic commerce applications \cite{2}.

There are many algorithms for integer factorization \cite{3}. From
the trivial trial division to the classical Fermat's factorization
method \cite{4} and Euler's factoring method \cite{5} to the modern
algorithms, the quadratic sieve \cite{6} and the number field sieve
\cite{7}. In particular the number field sieve algorithm was used
in 1996 to factor a 512 bit integer \cite{8}, the lowest integer
length used in commercial RSA implementations. There have been several
other big integers factored over the course of the last decade. I
would like to point out that in those cases the feat was accomplished
with tremendous effort developing the software and a very considerable
investment in hardware \cite{9},\cite{10}.

In what follows I will expose how MapReduce, a distributed computational
framework, can be used for integer factorization. As an example I
will show an implementation of the quadratic sieve algorithm. I will
also compare in terms of performance and cost a conventional implementation
with the MapReduce implementation.

\section{MapReduce}

I claim no participation in the development of the MapReduce framework.
This section is basically a short extract of the original MapReduce
paper by Jeff Dean and Sanjay Ghemawat \cite{11}. MapReduce is a
programming model inspired in computational programming. Users can
specify two functions, \emph{map} and \emph{reduce}. The \emph{map}
function processes a series of (key, value) pairs, and outputs intermediate
(key, value) pairs. The system automatically orders and groups all
(key, value) pairs for a particular key, and passes them to the reduce
function. The reduce function receives a series of values for a single
key, and produces its output, which is sometimes a synthesis or aggregation
of the intermediate values.

The canonical example of a MapReduce computation is the construction
of an inverted index. Let's take a collection of documents $\mathit{\mathcal{D}}=\left\{ D_{0},D_{1},...,D_{N}\right\} $
which are composed of words $D_{0}=\left(d_{0,0},d_{0,1},...,d_{0,L_{0}}\right),D_{1}=\left(d_{1,0},d_{1,1},...,d_{1,L_{1}}\right)$
and so on. We define a map function the following way:

\[
map:(i,D_{i})\rightarrow\left\{ \left(d_{i,0},\left(i,0\right)\right),\left(d_{i,1},\left(i,1\right)\right),...,\left(d_{i,L_{i}},\left(i,L_{i}\right)\right)\right\} \]

that is, for a given document it processes each word in the document
and outputs an intermediate pair. The key is the word itself, and
the value is the location in the corpus, indicated as (document, position).
The reduce function is defined as:

\[
reduce:\left\{ \left(d,\left(i_{1},j_{1}\right)\right),...,\left(d,\left(i_{L},j_{L}\right)\right)\right\} \rightarrow\left(d,\left\{ \left(i_{1},j_{1}\right),...,\left(i_{L},j_{L}\right)\right\} \right)\]

For a collection of pairs with the same key (the same word), it outputs
a new pair, in which the key is the same, and the value is the aggregation
of the intermediate values. In this case, the set of locations (document
and position in the document) in which the word can be found in the
corpus.

The MapReduce implementation automatically takes care of the parallel
execution in a distributed system, data transmission, fault tolerance,
load balancing and many other aspects of a high performance parallel
computation. The MapReduce model escales seamlessly to thousands of
machines. It is used continously for a multitude of real world applications,
from machine learning to graph computations. And most importantly
the effort required to develop a high performance parallel application
with MapReduce is much lower than using other models, like for example
MPI \cite{12}.

\section{Quadratic Sieve}

The Quadratic Sieve algorithm was conceived by Carl Pomerance in 1981.
A detailed explanation of the algorithm can be found in \cite{13}.
Here we will just review the basic steps. Let $N$ be the integer
that we are trying to factor. We will attempt to find $a,b$ such
that: $N\mid\left(a^{2}-b^{2}\right)\Rightarrow N\mid\left(a+b\right)\left(a-b\right)$.
If $\left\{ \left(a+b,N\right),\left(a-b,N\right)\right\} \neq\left\{ 1,N\right\} $
then we will have a factorization of $N$.

Lets define:\[
Q\left(x\right)=x^{2}-N\]

if we find $x_{1},x_{2},...x_{K}$ such that $\prod_{i=1}^{K}Q\left(x_{i}\right)$
is a perfect square, then: \[
N\mid\prod_{i=1}^{K}Q\left(x_{i}\right)-\left(\prod_{i=1}^{K}x_{i}\right)^{2}=\prod_{i=1}^{K}\left(x_{i}^{2}-N\right)-x_{1}^{2}x_{2}^{2}...x_{K}^{2}\]

\subsection{Finding Squares}

Let's take a set of integers $x_{1},...,x_{L}$ which are $B$-smooth
(all $x_{i}$ factor completely into primes $\leq B$). One way to
look for $i_{1},i_{2},...,i_{M}$ such that $\prod_{j=1}^{M}x_{i_{j}}$
is a square is as follows. Let's denote $p_{i}$ the i-th prime number.
$\prod_{j=1}^{M}x_{i_{j}}=p_{j_{1}}^{a_{1}}p_{j_{2}}^{a_{2}}...p_{j_{L}}^{a_{L}}$
is a square if and only if $2\mid a_{k}$ for all $k$ $\Leftrightarrow a_{k}\equiv0\, mod\,(2)$.
For each $x_{i}$ we will obtain a vector $v^{i}=v\left(x_{i}\right)$
where $v_{j}^{i}=max\left\{ k:p_{j}^{k}\mid x_{i}\right\} \, mod\,\left(2\right)$.
That is, each component $j$ of $v^{i}$ is the exponent of $p_{j}$
in the factorization of $x_{i}$ modulo $2$. For example, for $B=4$:

\begin{eqnarray*}
x_{1}=6,v^{1}=\left(1,1,0,0\right)\\
x_{2}=45,v^{2}=\left(0,0,1,0\right)\\
x_{3}=75,v^{3}=\mbox{\ensuremath{\left(0,1,0,0\right)}}\end{eqnarray*}

It is immediate that:

\[
v\left(\prod_{j=1}^{M}x_{i_{j}}\right)=\sum_{j=1}^{M}v\left(x_{i_{j}}\right)\]

Then\[
\prod_{j=1}^{M}x_{i_{j}}\mbox{is a square}\Leftrightarrow v\left(\prod_{j=1}^{M}x_{i_{j}}\right)=\overrightarrow{0}\]

In conclussion, in order to find a subset of $x_{1},...,x_{L}$ which
is a perfect square, we just need to solve the linear system:\[
\left(\begin{array}{ccccccc}
v^{1} & \mid & v^{2} & \mid & \ldots & \mid & v^{L}\end{array}\right)\left(\begin{array}{c}
e_{1}\\
e_{2}\\
\vdots\\
e_{L}\end{array}\right)\equiv\overrightarrow{0}\, mod\,(2)\]

\subsection{Sieving for smooth numbers}

Back to the original problem, we just need to find a convenient set
$\left\{ x_{1},x_{2},...,x_{L}\right\} $ such that $\left\{ Q\left(x_{1}\right),Q\left(x_{2}\right),...,Q\left(x_{L}\right)\right\} $
are $B$-smooth numbers for a particular $B$. First of all, lets
notice that we don't need to consider every prime number $\leq B$.
If a prime $p$ verifies: $p\mid Q(x)$ for some $x$ then:

\[
p\mid Q(x)\Leftrightarrow p\mid x^{2}-N\Leftrightarrow x^{2}\equiv N\, mod\,(p)\Leftrightarrow\left(\frac{N}{p}\right)=1\]

Because $N$ is a quadratic residue modulo p if and only if the Legendre
symbol of n over p is 1. We will take a set of primes which verifies
that property and we will call it \emph{factor base}.

In order to consider smaller values of $Q(x)$ we will take values
of $x$ around $\sqrt{N},$ i.e. $x\in\left[\lfloor\sqrt{N}\rfloor-M,\lfloor\sqrt{N}\rfloor+M\right]$
for some $M.$ Both $B$ above and $M$ here are chosen as indicated
in \cite{13}.

In order to factor all the $Q(x_{i})$ we will use a method called
\emph{sieving} which is what gives the quadratic sieve its name. Notice
that $p\mid Q(x)\Rightarrow p\mid Q(x+kp)=x^{2}+2kpx+k^{2}p^{2}-N=\left(x^{2}-N\right)+p\left(2kx+k^{2}p\right)$.
Then\[
Q(x)\equiv0\, mod\,(p)\Rightarrow\forall k\in\mathbb{N},Q(x+kp)\equiv0\, mod\,(p)\]

We can solve the equation $Q(x)\equiv0\, mod\,(p)\Leftrightarrow x^{2}-N\equiv0\, mod\,(p)$
efficiently and obtain two solutions $s_{1},s_{2}$ \cite{14}. If
we take: \[
z_{p,\left\{ 1,2\right\} }=min\left\{ x\in\left[\lfloor\sqrt{N}\rfloor-M,\lfloor\sqrt{N}\rfloor+M\right]:x\equiv s_{\left\{ 1,2\right\} }\, mod\,(p)\right\} \]
then all $Q\left(z_{p,\{1,2\}}+kp\right),k\in\left[0,K\right]$ are
divisible by $p$. We can divide each one of them by the highest power
of $p$ possible. For example:

\begin{eqnarray*}
\left(x_{i}\right)= & \left(\ldots,6,7,8,9,10,\ldots\right)\\
\left(Q\left(x_{i}\right)\right)= & \left(\ldots,-41,-28,-13,4,23,\ldots\right)\\
 & \left(\frac{77}{2}\right)=1\mbox{ as }77\equiv1\equiv1^{2}\, mod\,(2)\\
 & x^{2}-77\equiv0\, mod\,(2)\mbox{ yields }1,3,5,7,9,...\\
 & \left(\ldots,-41,-7,-13,1,23,\ldots\right) & \mbox{after sieving by }2\end{eqnarray*}

After sieving for every appropriate $p$, all the $Q(z)$ that are
equal to $1$ are smooth over the factor base.

\section{Method}

I developed a basic implementation of the Quadratic Sieve MapReduce
which runs on Hadoop \cite{15}. Hadoop is an open source implementation
of the MapReduce framework. It is made in Java and it has been used
effectively in configurations ranging from one to a few thousand computers.
It is also available as a commercial cloud service \cite{16}.

This implementation is simply a proof of concept. It relies too heavily
on the MapReduce framework and it is severy bound by IO. However the
size and complexity of the implementation are several orders of manitude
lower than many competing alternatives.

The 3 parts of the program are :
\begin{itemize}
\item \emph{Controller}: Is the master job executed by the platform. It
runs before spawning any worker job. It has two basic functions: first
it generates the factor base. The factor base is serialized and passed
to the workers as a counter. Second it generates the full interval
to sieve. All the data is stored in a single file in the distributed
Hadoop file system \cite{17}. It then relies on the MapReduce framework
to automatically split it in an adequate number of shards and distribute
it to the workers
\item \emph{Mapper}: The mappers perform the sieve. Each one of them receives
an interval to sieve, and they return a subset of the elements in
that input sieve which are smooth over the factor base. All output
elements of all mappers share the same key
\item \emph{Reducer}: The reducer receives the set of smooth numbers and
attempts to find a subset of them whose product is a square by solving
the system modulo 2 using direct bit manipulation. If it finds a suitable
subset, it tries to factor the original number, N. In general there
will be many subsets to choose from. In case that the factorization
is not succesful with one of them, it proceeds to use another one.
The single output is the factorization
\end{itemize}
In order to compare performance I developed another implementations
of the Quadratic Sieve algorithm in Maple. Both implementations are
basic in the sense that they implement the basic algorithm described
above and the code has not been heavily optimized for performance.
There are many differences between the two frameworks used that could
impact performance. Because of that a direct comparison of running
times or memory space may not be meaningful. However it is interesting
to notice how each of the implementations scales depending on the
size of the problem. The source code is available online at http://www.javiertordable.com/research.

\section{Results}

Figures 1 and 2 show the results both in absolute terms and normalized.
Figure 3 shows the disk usage of the MapReduce implementation. To
test both implementations I took a set of numbers of different sizes%
\footnote{1164656837, 117375210056563, 10446257742110057983, 1100472550655106750000029%
}. The number of decimal digits $d$ is indicated in the first column
of each table. In order to contruct those numbers I took two factors
close to $10^{\frac{d}{2}}$, with their product slightly over $10^{d}$. 

In each table sieve size indicates the number of elements that the
algorithm analyzed in the sieve phase. For the MapReduce application
the time result is taken from the logs, and the memory result is obtained
as the maximum memory used by the process. For the Maple implementation
both time and memory data are taken from the on screen information
in the Maple environment. Finally disk usage data for the MapReduce
is taken as the size of the file that contains the list of numbers
to sieve. The Maple program runs completely in memory for the samples
analyzed.

\begin{table}
\begin{centering}
\begin{tabular}{|c|c|c|c|c|c|}
\hline 
Decimal & Sieve & \multicolumn{2}{c|}{MapReduce} & \multicolumn{2}{c|}{Maple}\tabularnewline
\hline 
Digits & Size & Time ($s$) & Memory ($MB$) & Time ($s$) & Memory ($MB$)\tabularnewline
\hline
\hline 
$10$ & $5832$ & $2.0$ & $149.6$ & $0.1$ & $7.5$\tabularnewline
\hline 
$15$ & $85184$ & $3.0$ & $397.1$ & $3.5$ & $15.5$\tabularnewline
\hline 
$20$ & $970299$ & $35.0$ & $463.1$ & $116.0$ & $100.8$\tabularnewline
\hline 
$25$ & $7529536$ & $495.0$ & $670.0$ & $3413.7$ & $894.0$\tabularnewline
\hline
\end{tabular}
\par\end{centering}

~

\caption{Absolute performance of the MapReduce and Maple implementations}

\end{table}

\begin{table}
\begin{centering}
\begin{tabular}{|c|c|c|c|c|c|}
\hline 
Decimal & Sieve & \multicolumn{2}{c|}{MapReduce} & \multicolumn{2}{c|}{Maple}\tabularnewline
\hline 
Digits & Size & Time & Memory & Time & Memory\tabularnewline
\hline
\hline 
$10$ & $1.0$ & $1.0$ & $1.0$ & $1.0$ & $1.0$\tabularnewline
\hline 
$15$ & $14.6$ & $1.5$ & $2.7$ & $35.0$ & $2.1$\tabularnewline
\hline 
$20$ & $166.4$ & $17.5$ & $3.1$ & $1160.0$ & $13.4$\tabularnewline
\hline 
$25$ & $1291.1$ & $247.5$ & $4.5$ & $34137.0$ & $119.2$\tabularnewline
\hline
\end{tabular}
\par\end{centering}

~

\caption{Normalized performance of the MapReduce and Maple implementations}

\end{table}

\begin{table}
\begin{centering}
\begin{tabular}{|c|c|c|c|c|}
\hline 
Decimal & Absolute Sieve & Relative Sieve & Absolute & Relative\tabularnewline
\hline
\hline 
Digits & Size & Size & Disk ($MB$) & Disk ($MB$)\tabularnewline
\hline 
$10$ & $5832$ & $1.0$ & $0.1$ & $1.0$\tabularnewline
\hline 
$15$ & $85184$ & $14.6$ & $2.1$ & $14.6$\tabularnewline
\hline 
$20$ & $970299$ & $166.4$ & $29.4$ & $166.4$\tabularnewline
\hline 
$25$ & $7529536$ & $1291.1$ & $275.3$ & $1291.1$\tabularnewline
\hline
\end{tabular}
\par\end{centering}

~

\caption{Disk usage of the MapReduce implementation}

\end{table}

\section{Discussion}

The MapReduce implementation has a relatively big setup cost in time
and memory when compared with an application in a conventional mathematical
environment. However it scales better with respect to the size of
the input data.

MapReduce is optimized to split and distribute data form disk. If
an application handles a significant volume of data, IO capacity and
performance can be a limiting factor. In our case disk usage is directly
proportional to the size of the sieve set, which grows exponentially
on the number of digits.

Both MapReduce and Maple implementations are similar in terms of development
effort. The Maple implementation seems more adequate for small-sized
problems while the MapReduce application is more efficient for medium-sized
problems. Also it will be easier to scale in order to solve harder
problems.

\appendix
\end{document}